# Prediction of Room-Temperature Superconductivity in Quasi-atomic $H_2$-Type Hydrides at High Pressure

Qiwen Jiang[a], Defang Duan[a],*, Hao Song[b], Zihan Zhang[a], Zihao Huo[a], Shuqing Jiang[c,a], Tian Cui[b,a],*, Yansun Yao[d]

[a] *Key Laboratory of Material Simulation Methods & Software of Ministry of Education and State Key Laboratory of Superhard Materials, College of Physics, Jilin University, Changchun 130012, China*

[b] *Institute of High Pressure Physics, School of Physical Science and Technology, Ningbo University, Ningbo, 315211, China*

[c] *Synergetic Extreme Condition User Facility, College of Physics, Jilin University, Changchun, Jilin 130012, China*

[d] *Department of Physics and Engineering Physics, University of Saskatchewan, Saskatoon, Saskatchewan S7N 5E2, Canada*

*Corresponding author: duandf@jlu.edu.cn (Defang Duan), cuitian@nbu.edu.cn (Tian Cui)

# Abstract

Achieving superconductivity at room temperature (RT) is a holy grail in physics. Recent discoveries on high-$T_c$ superconductivity in binary hydrides $H_3S$ and $LaH_{10}$ at high pressure have directed the search for RT superconductors to compress hydrides with conventional electron-phonon mechanisms. Here, we predict an exceptional family of superhydrides under high pressures, $MH_{12}$ ($M$ = Mg, Sc, Zr, Hf, Lu), all exhibiting RT superconductivity with calculated $T_c$s ranging from 313 to 398 K. In contrast to $H_3S$ and $LaH_{10}$, the hydrogen sublattice in $MH_{12}$ is arranged as quasi-atomic $H_2$ units. This unique configuration is closely associated with high $T_c$, attributed to the high electronic density of states derived from $H_2$ antibonding states at the Fermi level and the strong electron-phonon coupling related to the bending vibration of $H_2$ and H-$M$-H. Notably, $MgH_{12}$ and $ScH_{12}$ remain dynamically stable even at pressure below 100 GPa. Our findings offer crucial insights into achieving RT superconductivity and pave the way for innovative directions in experimental research.

It was predicted that solid hydrogen would become metallic in the atomic state under extreme compression [1]. Theoretically, metallic hydrogen has all the ingredients needed for a room-temperature (RT) superconductor, *e.g.*, high-frequency phonons, high density of states (DOS) at the Fermi level, and very strong electron-phonon coupling (EPC) [2]. However, the direct compression of solid hydrogen may require pressure over 500 GPa [3-5] to reach a metallic state, which poses extreme difficulty in experiments. Alternatively, in compressed hydrides, the hydrogen species are 'precompressed' [6] by the metal elements, and therefore, the charge density sufficient for metallization can be achieved at less physical compression [7-10]. Currently, all experimentally synthesized high-$T_c$ hydrides necessitate high pressures above 150 GPa, but $T_c$ below RT [11-19]. The primary goal in physics and material science is to achieve RT superconductivity at pressures below 100 GPa, ultimately reaching ambient pressure conditions.

Diatomic $H_2$ is a ubiquitous building block of solid hydrogen (e.g., phase I, II, III [20-22]) and of many hydrides as well [7], such as theoretically predicted $SiH_4(H_2)_2$ [23] and $TeH_4$ [24], and experimentally synthesized $BaH_{12}$ [25], $SrH_{22}$ [26], $HfH_{14}$ [27], and $SnH_{12}$ [28], etc. These 'molecular' hydrides usually exhibit moderate superconductivity with a $T_c$ below 120 K since the hydrogen electrons occupy bonding orbitals far below the Fermi level, resulting in a low DOS, even an insulating state. Breaking $H_2$ molecular units is a prerequisite for achieving high-$T_c$ superconductivity [29-32]. To increase the $T_c$ in hydrides, one needs to 'free' electrons from their locked positions, that is, to weaken the $H_2$ units toward atomic hydrogen. This will allow the H electrons to occupy energy states closer to the Fermi level. Two prototypic hydrides consisting of atomic hydrogen sublattices, the cubic $H_3S$ [11,12,33,34] and cagelike $LaH_{10}$ [13-15,30,35,36], were predicted following this principle and subsequently confirmed by experiments. These two hydrides have very high measured $T_c$s in the 200-260 K range. Thus, the high-$T_c$ hydrides are recognized as being associated with weakened $H_2$ units.

In the present work, we have predicted a novel group of dodecahydrides, $MH_{12}$ ($M =$

Mg, Sc, Zr, Hf, Lu), calculated to have $T_c$ values exceeding RT. Different from previously reported atomic ($H_3S$), cage ($LaH_{10}$), and planar ($HfH_{10}$ [37]) forms, hydrogen atoms in $M$$H_{12}$ are regularly arranged in quasi-atomic peanut-like $H_2$ units. All dodecahydrides in this study are predicted to be RT superconductors with $T_c$ up to 313-398 K estimated by solving Eliashberg equations. In particular, $ScH_{12}$ and $MgH_{12}$ are the first RT hydride superconductors hitherto predicted to exist at pressures below 100 GPa. Adding one H atom into the $M$$H_{12}$, we obtained a class of $M$$H_{13}$ containing quasi-molecular $H_2$ units, which also exhibit high $T_c$s but are notably lower than their respective $M$$H_{12}$ counterparts. Our research provides a promising approach for investigating phonon-mediated RT superconductivity at lower pressures that meet or surpass the atomic metallic hydrogen.

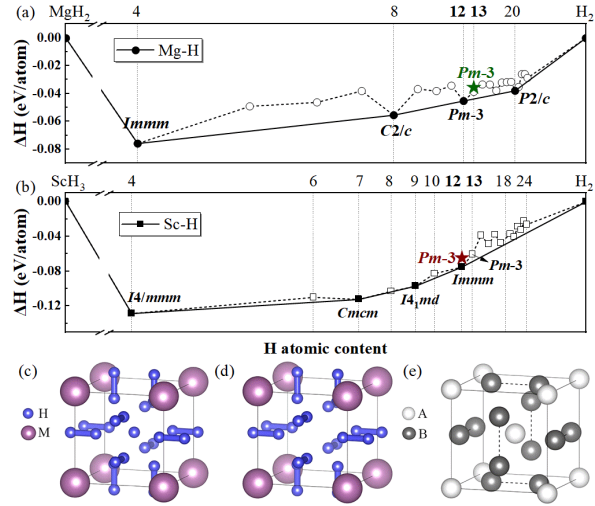

Fig. 1. The thermodynamic convex hull diagram of (a) Mg-H and (b) Sc-H systems at 300 GPa with ZPE included. The red and green stars near the convex hull correspond to the $Pm\bar{3}$-$ScH_{12}$ and $Pm\bar{3}$-$MgH_{13}$ phases, respectively. Crystal structure of (c) $M$$H_{13}$, (d) $M$$H_{12}$, and (e) the $A$15 type structure ($AB_3$, $Pm\bar{3}n$).

Recently reported high-$T_c$ hydrides mainly have the metal elements in the 2nd and 3rd groups of the periodic table [38]. Mg and Sc represent the lightest elements among them (except for Be) and serve as the primary focus of our study. We performed variable composition structural searches for Mg-H and Sc-H systems at 200 and 300 GPa using

AIRSS code [39]. The convex hull diagram in Fig. 1 illustrates the formation enthalpies for the optimal structures in all studied Mg-H and Sc-H stoichiometries. Zero-point energy (ZPE) was included in the enthalpy calculation to account for the quantum effects. In this search, we successfully recovered the previously predicted structures from Refs. [40-42]. Moreover, two unique hydrides, $M$H$_{12}$ and $M$H$_{13}$ ($M$ = Mg and Sc), with distinct $Pm\bar{3}$ symmetry, appeal to our attention. Both at 200 GPa and 300 GPa, MgH$_{12}$ and ScH$_{12}$ lies on the convex hull, while ScH$_{13}$ and MgH$_{13}$ slightly surpass the hull by about ~8 meV/atom, as shown in Fig. 1 and Fig. S1 of supplementary material (SM). Due to synthesizing hydrides typically involving laser heating [18,19,43-45], we considered the temperature effect through quasi-harmonic free energy calculations (Fig. S2). We found that the $Pm\bar{3}$ phase of MgH$_{12}$ is more stable than the $R$-3 phase at above 200 GPa and 1400 K, while the $Pm\bar{3}$ phase of ScH$_{12}$ becomes more stable than the ground state $P6_4$ and $Immm$ at temperatures above 1700 K within the range of 200-400 GPa. These results suggest the synthesis potential of ScH$_{12}$ and MgH$_{12}$ within the pressure range of 200-300 GPa, alongside high temperatures.

To examine the dynamical stability of $M$H$_{12}$ and $M$H$_{13}$ ($M$ = Mg and Sc), we calculated their phonon spectra at different pressures (Figs. S21-24). There are no imaginary phonon frequencies for MgH$_{12}$ at 210 GPa and above, and for MgH$_{13}$ at 300 GPa and above, which establish their dynamic stability ranges. ScH$_{12}$ and ScH$_{13}$ are dynamically stable down to a low pressure of 90 GPa. We extended the investigation to other isostructural $M$H$_{12}$ and $M$H$_{13}$ compounds involving other metal elements and identified Zr, Hf, and Lu, guided by criteria of similar Pauling electronegativity (~1.3) and atomic radius (~1.6 Å). Subsequent phonon dispersion relation calculations confirm that these three metals' corresponding $M$H$_{12}$ and $M$H$_{13}$ are dynamically stable (Fig. S22-24). In comparison, $M$H$_{12}$ and $M$H$_{13}$ hydrides with Ca, Y, La, Ac and Th are dynamically unstable up to 600 GPa due to large atomic radius or low electronegativity (Fig. S6).

In the $Pm\bar{3}$-$M$H$_{13}$, $M$ atoms form a cubic lattice with a single H atom occupying the center. A pair of H$_2$ units are perpendicular to and intersecting each face of the cube (Fig. 1c). If the centred H atom is removed, the structure will turn into the $Pm\bar{3}$-$M$H$_{12}$ structure (Fig. 1d). The structures of $M$H$_{12}$ and $M$H$_{13}$ are viewed as variants of the A15 type (AB$_3$, $Pm\bar{3}n$), with $M$ and diatomic H$_2$ being the two bases (see Fig. 1e). The structural information is listed in the Table S1. Interestingly, the $A$15 structure has been known to exist in conventional superconductors, such as Nb$_3$Sn and Nb$_3$Ge [46]. There are also $A$15-type superconductors in metal hydrides. For example, GaH$_3$ and GeH$_3$ have high $T_c$ of 123 K at 120 GPa [46] and 140 K at 180 GPa [47], respectively. Ternary hydride LiPH$_6$ [48] also adopts the $A$15-like configuration, which is predicted to be a high-$T_c$ superconductor (150-167 K at 200 GPa).

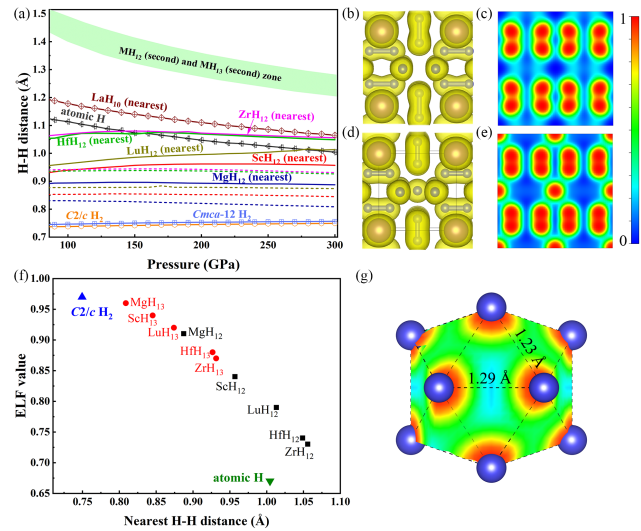

Fig. 2. (a) Distance between H atoms in $M$H$_{12}$ (solid lines) and $M$H$_{13}$ (dashed lines of the same color) compared to atomic and molecular hydrogen as a function of pressure. 3D ELF of (b) ScH$_{12}$ and (d) ScH$_{13}$ with isosurface value of 0.6. 2D ELF of (c) ScH$_{12}$ and (e) ScH$_{13}$ for the (0 0 2) plane at 300 GPa. (f) The ELF value varies with the nearest H-H distance of $M$H$_{12,13}$, molecular H$_2$ and atomic H at 300 GPa. (g) ELF cut plane of hydrogen icosahedron of ScH$_{12}$ at 300 GPa.

The bonding characteristics of $M$H$_{12}$ and $M$H$_{13}$ are investigated using electron localization function (ELF), crystal orbital Hamiltonian population (COHP), and Bader charge analysis. Bader analysis (Table S2) reveals significant charge transfer from the $M$ atom to the H atoms, while each H$_2$ pair receives more electrons in $M$H$_{12}$ than in $M$H$_{13}$. ScH$_{12,13}$ serves as a typical example at 300 GPa. In ScH$_{12}$, each H$_2$ pair accepts 0.190 e⁻, while each H$_2$ pair accepts 0.164 e⁻ in ScH$_{13}$. Since the electrons acquired by hydrogen must populate their antibonding states, this weakens the H-H bonds. In Fig. 2a, we compare the nearest and the second nearest H-H distances of $M$H$_{12}$ and $M$H$_{13}$ to those in atomic H ($I4_1/amd$) and molecular H$_2$ solids ($C2/c$ and $Cmca$-12 phases). H-H contacts in both $M$H$_{12}$ and $M$H$_{13}$ are longer than the covalent bond length in molecular H$_2$, as one would expect for weakened bonds. On average, shorter H-H bonds in $M$H$_{13}$ than in $M$H$_{12}$ are consistent with fewer electrons in the antibonding orbitals. Interestingly, similar to molecular H$_2$, the H-H contact in $M$H$_{13}$ does not change with pressure, indicating that the latter still contains molecular H$_2$, albeit weaker in strength. We thus term the H$_2$ units in $M$H$_{13}$ as 'quasi molecular'. On the other hand, the H-H bond length in $M$H$_{12}$ keeps increasing at higher pressures, showing a pronounced tendency for H$_2$ molecule dissociation, and reaches almost the same value as that in atomic hydrogen at 300 GPa. The important molecular feature of H$_2$ molecular stretching vibrational modes near the frequency of 3000 cm$^{-1}$ is absent in $M$H$_{12}$ (see Fig. S21-22). Hence, it seems reasonable to refer to the H$_2$ unit in $M$H$_{12}$ as 'quasi atomic'. The electron clouds surrounding quasi-atomic H$_2$ are shaped like peanuts, with a high ELF value of ~0.84 for ScH$_{12}$ (Fig. 2b-2c and Figs. S6-S9). Additionally, the ELF value between adjacent H$_2$ units (perpendicular) at a distance of 1.23 Å is approximately 0.45, exhibiting metallic interaction with free electrons connectivity (Fig. 2g). As shown in Fig. 2f, the H-H bonds in $M$H$_{12}$ and $M$H$_{13}$ display distinct electron localization characteristics, with the ELF values showing an approximately linear correlation with the H-H bond lengths. The calculated integrated crystalline orbital Hamiltonian population (ICOHP) values within the H$_2$ units (Figs. S11-12) are -3.23 eV in ScH$_{12}$ and -4.91 eV in ScH$_{13}$, indicating a much stronger bonding interaction in the latter. The ICOHP values between the H$_2$ units (perpendicular) are -0.91 eV in ScH$_{12}$ and -0.55 eV in ScH$_{13}$. This manifests a weak inter-molecular interaction in ScH$_{12,}$ which causes the weakening of the H$_2$ units.

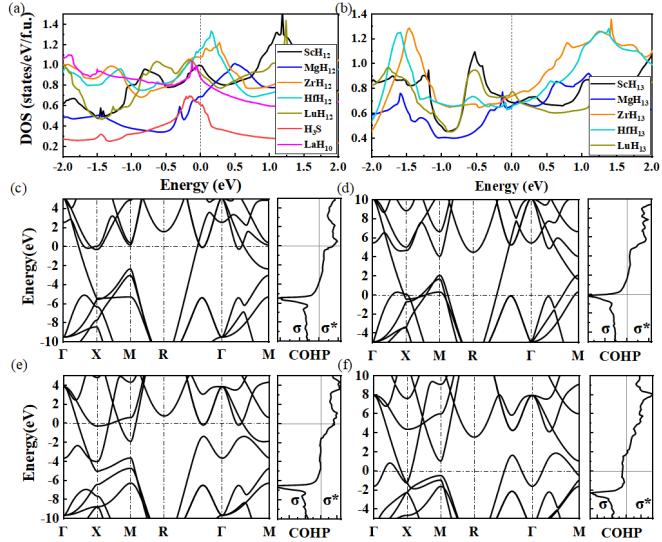

Fig. 3. Total DOS of (a) $M$H$_{12}$ and (b) $M$H$_{13}$ compared to H$_3$S and LaH$_{10}$. Electronic band structures and COHP of H-H bonds for (c) MgH$_{12}$, (d) Mg$_0$H$_{12}$, (e) MgH$_{13}$, and (f) Mg$_0$H$_{13}$ at 300 GPa.

The electronic band structures and project DOS of $M$H$_{12}$ and $M$H$_{13}$ are shown in Fig. 3 and Figs. S27-S31. Given the uniqueness of magnesium superhydrides, we first discuss other $X$H$_{12, 13}$ ($X$ = Sc, Zr, Hf, Lu). A common feature displayed in these $X$H$_{12}$ is the van Hove singularity with a large total DOS value at the Fermi level (Fig. 3a), formed by strong hybridization of $d$-orbital ($f$-orbital) of metal atoms and the $s$-orbital of H atoms (Figs. S28-S31). In contrast, the DOS in $X$H$_{13}$ is notably uniform around the Fermi level (Fig. 3b). The van Hove singularity plays an important role in enhancing the EPC and increasing $T_c$ in H$_3$S and LaH$_{10}$. The comparison of $X$H$_{12}$ to H$_3$S and LaH$_{10}$ at 300 GPa shows that $X$H$_{12}$ has a significantly larger DOS at the Fermi level, with a trend of $X$H$_{12}$ > LaH$_{10}$ > $X$H$_{13}$ > H$_3$S (Fig. 3a). High electronic DOS at the Fermi level, in particular those induced by hydrogen, sets a favourable condition for Cooper pairs, strong EPC and superconductivity. The electronic structures of MgH$_{12}$ and MgH$_{13}$ are different from $X$H$_{12, 13}$, characterized by nearly unoccupied $d$-states of Mg in the valence bands, and H atoms significantly contribute to the total DOS (~86%). As shown in Figs. 3c-3f, the band structures of MgH$_{12}$ and MgH$_{13}$ highly resemble Mg$_0$H$_{12}$ and Mg$_0$H$_{13}$ (hypothetical structures with Mg removed from the hydrides, without relaxation), exhibiting a pure hydrogen character. The presence of Mg provides electrons to the $H_2$ units, which elevate the Fermi level from the low-energy bonding states (σ) to the antibonding states ($σ^*$), maximizing the electronic states of H at the Fermi level. We observe interlocking electron pockets near the X point, especially in $MgH_{12}$, where the nearly degenerate $A_g$ and $A_u$ bands at the X point and the hole pocket at the Γ point, create a fish-tail-shaped Fermi surface along the Γ-X direction (Fig. S30). The subsequent peak of the nesting function ξ(Q) in that direction indicates strong Fermi surface nesting, which is advantageous for EPC (Fig. S31).

To examine the superconducting properties of $MH_{12,13}$, we calculated the logarithmic average phonon frequency ($ω_{log}$), EPC parameter (λ), and Eliashberg phonon spectral function [$α^2F(ω)$] (Figs. S21-23). It should be noted that the vibrational bands of hydrogen in $MgH_{12}$ and $ScH_{12}$ hydrides are mixed and dispersive, with all frequencies below 2800 $cm^{-1}$ (Fig. S21) at 300 GPa, consistent with the disappearance of molecular $H_2$. For $MgH_{13}$ and $ScH_{13}$, the high-frequency vibrations still form an isolated non-dispersive block, peaking at approximately 3000 $cm^{-1}$. This frequency aligns with the stretching vibrations of quasi-molecular $H_2$, yet their contribution to the EPC is minor (< 10%). For pure molecular hydrogen *C2/c* and *Cmca*-12 phases [49], the highest vibrational frequency is around 4000 $cm^{-1}$ at 250 GPa. The lower hydride vibrational frequencies indicate weaker H-H bonds induced by the interaction between H and metal atoms. Compared to $MH_{13}$, the calculated λ values for all $MH_{12}$ compounds within their dynamically stable pressure range are greater than or equal to 2.5, suggesting a very strong EPC. The remarkable λ values are mainly associated with mixed stretching, rocking and scissoring vibrations of the $H_2$ units within the mid-frequency range, accompanied by the bending vibrations of H-Sc-H (Fig. S25). Using typical Coulomb pseudopotential $μ^*$ =0.1-0.13, applying the Eliashberg equations yields a very high $T_c$ value ranging 304-325 K for $ScH_{12}$ with λ = 2.50 at 300 GPa (Table S3). In the case of $ScH_{13}$, the highest $T_c$ is calculated to be 185 K ($μ^*$ = 0.1) with λ = 1.40, much lower than that of $ScH_{12}$. $MgH_{12}$ and $MgH_{13}$ exhibit RT superconductivity, with $T_c$ values of 366-388 K and 303-324 K at 300 GPa, respectively. $MH_{12}$ (*M*=Zr, Hf, and Lu) are also calculated to have RT $T_c$ between 333 K and 388 K, and $M$H$_{13}$ hosts high $T_c$ with 160-200 K ($\mu^* = 0.10$).

As shown in Fig. 4a, the superhydrides $M$H$_{12}$ investigated in this study exhibit remarkably high $T_c$ values comparable to atomic metallic hydrogen (~356 K at 500 GPa [50]), particularly with substantially reduced stability pressures in Sc and Mg. This highlights the potential of $M$H$_{12}$ superhydrides as promising candidates for RT superconductors. Strikingly, the superconductivity of ScH$_{12}$ shows weak dependence on pressure, maintaining RT superconductivity down to a significantly lower pressure of 90 GPa when compared to other compressed hydrides (e.g., YH$_{10}$ and Li$_2$MgH$_{16}$ at 250 GPa).

Given the light mass of hydrogen, anharmonicity is crucial for superhydrides, especially those with strong EPC [51,52]. Due to the substantial computational cost, we take MgH$_{12}$ as an example and examined this effect using the stochastic self-consistent harmonic approximation [53]. As shown in Fig. S26, the anharmonic correction hardens phonon modes near the Γ-point, eliminating instability above 70 GPa. At 210 GPa, the anharmonicity weakens the EPC parameter while enhancing $\omega_{\log}$, resulting $T_c$ slightly decreases to 345-365 K (Table S1), maintaining above RT. At 70 GPa, the anharmonic $T_c$ value is 281-296 K, still close to RT.

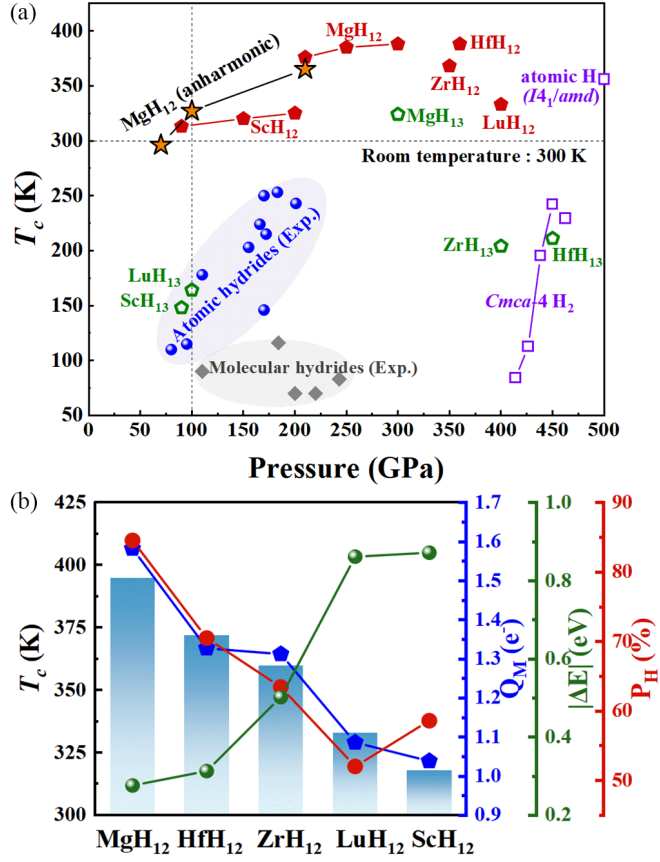

Fig. 4. (a) Calculated $T_c$ of $M$H$_{12}$ and $M$H$_{13}$ by numerically solving Eliashberg equations at different pressures with $\mu^* = 0.10$. The blue circles and gray squares denote the well-known hydrides from experiments [11-19,27,28,44,45,54-60]. Superconductivity in molecular hydrogen ($Cmca$-4) and atomic hydrogen ($I4_1/amd$) is documented in Ref. [50,61]. (b) Calculated at 400 GPa for $M$H$_{12}$, the $T_c$, the electrons transferred from $M$ to hydrogen atoms ($Q_M$), the percentage of the H electronic DOS at the Fermi level ($P_H$), and the distance between the H-H antibonding state and the Fermi level at the X point ($|\Delta E|$).

As discussed above, the intriguing RT superconductivity of $M$H$_{12}$ is related to the transferred electrons from metal $M$ to H atoms ($Q_M$), the percentage of H electronic DOS ($P_H$), and the energy difference ($|\Delta E|$) between the flat band at the X point and the Fermi level (Fig. S34). To gain further insights, we plotted the relationship between these parameters and $T_c$, as shown in Fig. 4b. From Sc to Lu, Zr, Hf, and finally to Mg, there is an increasing electron transfer from metals to hydrogen, elevating the Fermi energy and leading to a decrease in $|\Delta E|$. This brings the antibonding orbitals of hydrogen corresponding to the flat band closer to the Fermi level, maximizing the DOS contribution of hydrogen and promoting the formation of Cooper pairs. Consequently, the EPC constant for the optical modes derived from hydrogen at the X point gradually strengthens (Fig. S35), increasing $T_c$.

In previous research, it was thought that hydrides containing $H_2$ or $H_3$ molecular units are unfavorable for superconductivity. However, this study found a particular case, $MgH_{13}$, with quasi-molecular $H_2$ units exhibiting RT superconductivity. As shown in Fig. S36, the H-H bond length does not display a singular linear relationship with $T_c$. Hence, inferring $T_c$ based solely on this criterion appears to be rudimentary. We further performed a statistical analysis of the relationship between $T_c$ and the product of $P_H$ and the DOS per hydrogen atom ($DOS_{per\ H}$), and the results show a positive correlation (Fig. S37). It is necessary to calculate the DOS of hydrogen at the Fermi level to predict the possibility of hydride superconductivity [62]. The pure hydrogen framework of $H_{12}$ offers a plethora of unoccupied energy levels. In certain s/p block metal hydrides like Mg, extra electrons support the prominent DOS associated with the dominant antibonding states of $H_2$, albeit with a minor inclination towards molecular dissociation.

In summary, an extensive structure search combined with *ab initio* calculations reveals a unique superhydride $MH_{12}$ ($M$ = Mg, Sc, Zr, Hf, Lu) under high pressure. In this structure, hydrogen atoms are arranged in quasi-atomic peanut-like $H_2$ units, which differs significantly from the hydrogen configurations in known high-$T_c$ hydrides $H_3S$ and $LaH_{10}$. $MH_{12}$ is calculated to have a superconducting critical temperature $T_c$ similar to or superior to metallic hydrogen solids, exhibiting superconductivity at the highest temperature among all known/predicted binary hydrides. In $MH_{13}$, hydrogen atoms form quasi-molecular $H_2$ units, which are also predicted to possess high $T_c$. The intriguing family of RT superconducting hydrides we have predicted is poised to stimulate future high-pressure experimental investigations, potentially enabling synthesis within diamond anvil cells through laser heating.

## Acknowledgements

This work was supported by the National Key Research and Development Program of China (No. 2022YFA1402304), the National Natural Science Foundation of China (Grants No. 12122405, No. 52072188, No. 12274169, and No. 51632002), the Natural Sciences and Engineering Research Council of Canada (NSERC), program for Science and Technology Innovation Team in Zhejiang (No. 2021R01004) and the Fundamental Research Funds for the Central Universities. Some calculations were performed at the High Performance Computing Center of Jilin University and using TianHe-1(A) at the National Supercomputer Center in Tianjin.

## Competing interests

The authors declare no competing interests.